\documentclass[paper]{aa}

\usepackage{graphicx}

\newcommand{\beq}{\begin{equation}}
\newcommand{\eeq}{\end{equation}}
\newcommand{\beqa}{\begin{eqnarray}}
\newcommand{\eeqa}{\end{eqnarray}}

\begin{document}

\title
{
The excitation of $5$-min oscillations in the solar corona
}

\subtitle{}

\author{
      T.V. Zaqarashvili,  \inst{1,4}
      K. Murawski,      \inst{2}
      M.K. Khodachenko,   \inst{1}
      D. Lee   \inst{3}
       }
\offprints{T. Zaqarashvili \email{teimuraz.zaqarashvili@oeaw.ac.at}}
\institute{
Space Research Institute, Austrian Academy of Sciences, Schmiedlstrasse 6, 8042 Graz, Austria\\
  \email{teimuraz.zaqarashvili@oeaw.ac.at, maxim.khodachenko@oeaw.ac.at}
  \and
Group of Astrophysics,
Institute of Physics, UMCS, ul. Radziszewskiego 10, 20-031 Lublin, Poland\\
  \email{kmur@kft.umcs.lublin.pl}
    \and
ASC/Flash Center, The University of Chicago, 5640 S. Ellis Ave, Chicago, IL 60637, USA\\
\email{dongwook@flash.uchicago.edu}
\and
Abastumani Astrophysical Observatory at Ilia State University, Kazbegi ave. 2a, Tbilisi, Georgia
            }
\date{received / accepted }

\abstract
{}
{
We aim to study excitation of the observed $\sim$ $5$-min oscillations in the solar corona by
localized pulses that are launched in the photosphere.
} {We solve the full set of nonlinear one-dimensional
Euler equations numerically for the velocity pulse propagating in
the solar atmosphere that is determined by the realistic temperature
profile. } {Numerical simulations show that an initial velocity pulse
quickly steepens into a leading shock,
while the nonlinear wake in the chromosphere leads to the formation
of consecutive pulses. The time interval between arrivals of two
neighboring pulses to a detection point in the corona is
approximately $5$ min. Therefore, the consecutive pulses may result
in the $\sim$ $5$-min oscillations that are observed in the solar
corona.} {The $\sim$ 5-min oscillations observed in the solar
corona can be explained in terms of consecutive shocks that result
from impulsive triggers
launched within the solar photosphere by granulation and/or reconnection.
}
\titlerunning{$5$-min oscillations in the solar corona}
\authorrunning{T.V. Zaqarashvili et al.}

\maketitle
%
%
%
\section{Introduction}
Propagating acoustic waves are frequently seen in the solar corona
as periodic variations of spectral line intensity (De Moortel et al.
\cite{de Moortel00,de Moortel02}, Marsh et al. \cite{marsh03}, Lin
et al. \cite{lin2005,lin2006}, Srivastava et al.
\cite{srivastava08}, Wang et al. \cite{wang09}). As these waves are
often observed within the frequency range corresponding to the
acoustic waves in the solar photosphere/chromosphere, this logically
leads to the idea of penetration of the photospheric acoustic
oscillations into the corona. However, the photospheric $5$-min
oscillations are evanescent in the gravitationally stratified solar
atmosphere as their frequency is lower than the cut-off frequency
(Lamb \cite{lamb1908}, Roberts \cite{roberts}, Musielak et al.
\cite{musielak2006}). Bel \& Leroy (\cite{bel77}) suggested that the
cut-off frequency of the magnetic field-free atmosphere is lower for
waves propagating obliquely to the vertical direction. De Pontieu et
al. (\cite{dep05}) proposed that p-modes may be channeled into the
solar corona along inclined magnetic field lines as a result of the
decrease of the acoustic cut-off frequency. McIntosh \& Jefferies
(\cite{mcintosh06}) found the observational justification of the
modification of the cut-off frequency by inclined magnetic field. As
the magnetic field of active region loops is predominantly vertical
in the photosphere/chromosphere, it is unclear how p-modes may
penetrate into the coronal regions.

There are two different types of drivers in the highly dynamic solar
photosphere: oscillatory (e.g. p-modes) and impulsive (e.g.
granulation and/or explosive events due to magnetic reconnection).
Both types of drivers may be responsible for the observed dynamical
phenomena in upper atmospheric regions.

As a result of the rapid decrease of mass density
with height, finite-amplitude
high-frequency
photospheric oscillations can quickly grow in their amplitudes and steepen into shocks, which
by energy dissipation can lead to the chromospheric heating (Narain \&
Ulmschneider \cite{narain,narain96}, Carlsson \& Stein
\cite{carlson97}, Ruderman \cite{ruderman}). Lower-frequency waves,
those with $\sim$5 min period, are not good candidates for the
chromospheric heating (Narain \&  Ulmschneider \cite{narain}).

It was found by Hollweg (\cite{hol82}) that a localized pulse that is launched initially sets up a nonlinear wake
which results in a trail of consecutive shocks. Such shocks were called rebound shocks by Hollweg (\cite{hol82}).

The time interval between consecutive shocks is close to the period of the nonlinear wake.
In the linear case, the wake oscillates with the acoustic cut-off frequency of the stratified atmosphere.
Nonlinearity modifies the wave period of the wake,
with many features
of spicules
which exhibit the periodicity of
about $5$-min (Murawski \& Zaqarashvili \cite{murawski2010}).
Then, these quasi-periodic shocks may lead to the oscillatory dynamics of coronal plasma, which is observed
as intensity oscillations in coronal spectral lines.

The aim of this paper is to study the role of rebound shocks,
which are formed by an impulsive perturbation, on
the observed $\sim 5$-min oscillations in the solar corona. Here we
consider the simplest hydrodynamic case, which can be developed to
more realistic magnetohydrodynamic model in future studies.

This paper is organized as follows. The basic equations and the
atmospheric model are described in Sect.~\ref{sect:num_model}. The
numerical model and results of numerical simulations of impulsive
photospheric driver are discussed in Sect.~\ref{sec:num_res}. This
paper is concluded by a summary of the main results in
Sect.~\ref{sec:sum}.
\section{Basic model}\label{sect:num_model}
\subsection{
Hydrodynamic equations
} \label{sect:equ_model}
Our model system is taken to be composed of a gravitationally-stratified solar atmosphere
that is described by
one-dimensional (1D)
Euler
equations:
\beqa
\label{eq:MHD_rho}
\frac{\partial \varrho}{\partial t}+\frac{\partial (\varrho V)}{\partial y} = 0\, ,\\
\label{eq:MHD_V}
\varrho{{\partial V}\over {\partial t}}+ \varrho V \frac{\partial V}{\partial y}=-\frac{\partial p}{\partial y} - \varrho g\, , \\
\label{eq:MHD_p}
{\partial p\over \partial t} + \frac{\partial (pV)}{\partial y} = (1-\gamma)p \frac{\partial V}{\partial y}\, .
\eeqa
Here ${\varrho}$ denotes the mass density, $V$ is a vertical component of the flow velocity,
$p = {k_{\rm B}} \varrho T / {m}$ is the gas pressure, $\gamma=5/3$ is the adiabatic index,
$g=272$ m s$^{-2}$ is the gravitational acceleration, $T$ is the
temperature, $m$ is the mean particle mass and $k_{\rm B}$ is
Boltzmann's constant.
\subsection {The equilibrium}
We assume that at the equilibrium the solar atmosphere is settled in a
static ($V = 0$) environment
in which
the pressure gradient force is balanced by the gravity, that is
\beq
-\frac{\partial p_{\rm 0}}{\partial y} - \varrho_{\rm 0} g = 0\;.
\label{eq:p}
\eeq
%
With the use of
the equation of state
we obtain the equilibrium gas pressure and mass density as
\beq
\label{eq:pres}
p_{\rm 0}(y) = p_{\rm 00}~\exp\left( - \int_{y_{\rm r}}^{y} \frac{dy^{'}}{\Lambda (y^{'})} \right ),\hspace{3mm}
\varrho_{\rm 0} (y) = \frac{p_{\rm 0}(y)}{g \Lambda (y)}\, .
\eeq
Here $\Lambda(y) = k_{\rm B} T(y)/(mg)$
%
%
is the pressure scale-height and $p_{\rm 00}$ is the gas pressure at
the reference level, chosen here at $y_{\rm r}=10$ Mm.

We adopt a realistic temperature profile $T_{\rm }(y)$ for the solar atmosphere (Vernazza et al. 1981).
In this profile $T_{\rm }$ attains a value of
about $5700$ K at the top of the photosphere which corresponds to $y=0.5$ Mm. At higher altitudes $T_{\rm }(y)$
falls off until
it reaches its minimum of $4350$ K at
$y\simeq 0.95$ Mm. Higher up $T_{\rm }(y)$ grows gradually with height
up to the transition region which is located at $y\simeq 2.7$ Mm. Here $T_{\rm }(y)$ experiences a sudden growth up to the coronal
value of $1.5$ MK at $y=10$ Mm.
Having specified $T_{\rm }(y)$ we can obtain mass density and gas pressure
profiles with the use of Eq.~(\ref{eq:pres}).

Equations~(\ref{eq:MHD_rho})-(\ref{eq:MHD_p})
are
to be solved numerically
for
an impulsive perturbation launched initially within the solar photosphere.
The numerical simulations for a harmonic driver were already reported elsewhere
(e.g. Erd\'elyi et al. \cite{erd2007}, Fedun et al. \cite{fedun09} and references therein).
Therefore,
it is
justifiable to limit our study to the case of impulsively generated waves.

\subsection{Linear approximation}

Neglecting all nonlinear terms in
Eqs.~(\ref{eq:MHD_rho})-(\ref{eq:MHD_p}) leads to the classical
Klein-Gordon equation (Rae and Roberts \cite{rae}, Roberts
\cite{roberts}) \beq \frac{\partial^2 Q}{\partial t^2} - c^2_{\rm
s}\frac{\partial^2 Q}{\partial y^2}+ \Omega^2_{\rm c} Q= 0,
\label{eq:klein} \eeq where $Q(t,y)=V(t,y)\sqrt{\varrho_{\rm 0}
(0)c^2_{\rm s}(0)/\varrho_{\rm 0} (y)c^2_{\rm s}(y)}$ and \beq
\Omega^2_{\rm c}(y)=\frac{c^2_{\rm s}}{4\Lambda}\left
(1+2\frac{\partial \Lambda}{\partial y} \right ). \label{eq:cut-off}
\eeq In the case of an isothermal atmosphere, the Klein-Gordon
equation implements a natural frequency of the stratified medium,
$\Omega_{\rm c}={c_{\rm s}/{2\Lambda}}$, called the cut-off
frequency for acoustic waves. A physical interpretation of the
cut-off frequency is that waves of lower (higher) frequencies than
$\Omega_{\rm c}$ are evanescent (propagating). According to the
theory developed for the linear Klein-Gordon equation the initially
launched pulse results in a leading pulse which propagates with the
sound speed. This pulse is followed by the wake, which oscillates
with
$\Omega_{\rm c}$ and gradually decays as time progresses (Lamb
\cite{lamb1908}, Rae and Roberts \cite{rae}). However, situation is
changed when one considers nonlinear terms in
Eqs.~(\ref{eq:MHD_rho})-(\ref{eq:MHD_p}). Then, a large-amplitude
initial pulse may lead to consecutive shocks due to nonlinearity
(Hollweg \cite{hol82}). In the nonlinear regime, the wake may
oscillate with the period, which differs from the linear cut-off
period. Then the interval between consecutive shocks may depend on
the amplitude of the initial pulse (Murawski and Zaqarashvili
\cite{murawski2010}). We solve the fully nonlinear equations for a
realistic atmospheric model numerically and the results are
presented in the next section.

\section{Numerical simulations
for impulsively generated waves
}\label{sec:num_res}

Equations~(\ref{eq:MHD_rho})-(\ref{eq:MHD_p}) are solved with a use of the code FLASH (Dubey et al. 2009).
We set the simulation region as $-0.5$ Mm $\le y\le 29.5$ Mm and at the numerical boundaries we fix in time
all plasma quantities to their equilibrium values.
In our studies we use an adaptive mesh refinement (AMR) grid with a
minimum (maximum) level of refinement blocks set to $3$ ($9$).
%
In the simulations, we excite waves by launching at $t=0$ s
the initial velocity pulse of a Gaussian profile

%
\beq
\label{eq:perturb}
V(y,t=0) = A_{\rm v}\exp\left[
                                      -\frac{(y-y_{\rm 0})^2}{w^2}
                                \right]\,
\eeq
with $A_{\rm v}$ the amplitude of the initial  pulse, $y_{\rm 0}$ its initial position and
$w$ its width. We set
and hold
fixed $y_{\rm 0}=0.5$ Mm and $w=0.1$ Mm, but allow $A_{\rm v}$ to vary in different simulations.

%
\begin{figure}
\begin{center}
{
\includegraphics[scale=0.5]{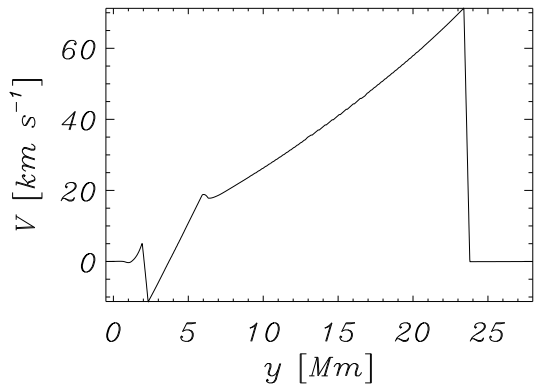}
}
\caption{
Velocity (in units of $1$ km s$^{-1}$) profile vs height
$y$ (in units of $1$ Mm) at $t=250$ s for $A_{\rm v} = 1$ km
s$^{-1}$. Here $y=0$ Mm ($y\simeq 2.7$ Mm) corresponds to the base
of the photosphere (the transition region).
}
\label{fig:spat_prof}
\end{center}
\end{figure}

At the initial stage of the wave evolution
the initial pulse splits into counter-propagating waves.
As a result of the rapid decrease of the equilibrium mass density,
the upward propagating pulse grows in its amplitude and
quickly steepens into a shock.
Figure~\ref{fig:spat_prof} illustrates the spatial profile of $V(y)$
at $t=250$ s
for
$A_{\rm v} = 1$ km s$^{-1}$. At this time, the pulse reaches the
level $y=23.5$ Mm, while the secondary shock begins to form from the
nonlinear wake at $y=2$ Mm. Later on, this secondary shock
propagates upwards, and then it is followed by the next shock. This
process repeats in time until the perturbation energy finally
subsides to zero.
%
\begin{figure}
\begin{center}
\includegraphics[scale=0.5]{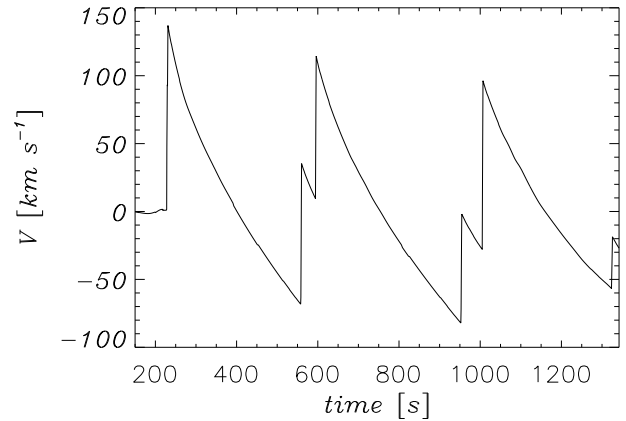}
\includegraphics[scale=0.5]{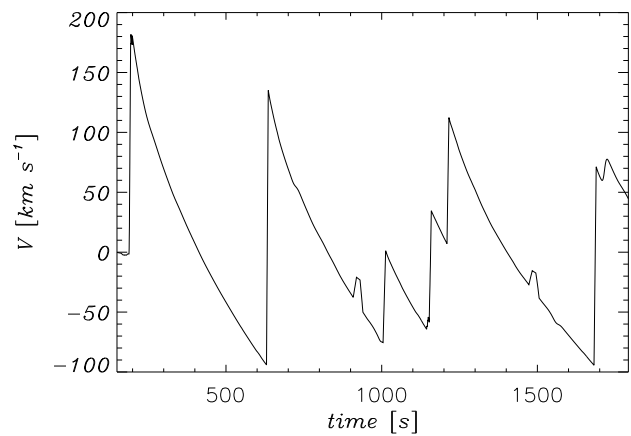}
\caption{\small Time-signatures of $V$ (in units of km s$^{-1}$)
collected at $y=20$ Mm for the case of $A_{\rm v}=1$ km s$^{-1}$
(top panel) and $A_{\rm v}=2$ km s$^{-1}$ (bottom panel).
Time is expressed in units of $1$ s.
}
\label{fig:time_sig}
\end{center}
\end{figure}

Figure~\ref{fig:time_sig} displays the temporal dynamics of the
velocity that is
collected at
$y=20$ Mm for the initial pulse amplitude of $A_{\rm v}=1$ km
s$^{-1}$ (top panel) and $A_{\rm v}=2$ km s$^{-1}$ (bottom panel).
In the top panel,
the arrival of the leading shock to the detection point occurs at $t\simeq 250$ s and
the second pulse reaches the detection point at $t\simeq 570$ s i.e. after $\sim 320$ s.
Thus, the nonlinear wake in the chromosphere excites quasi-periodic
upward propagating pulses, which enter the corona and may cause
quasi-periodic intensity oscillations which are observed in the
imaging data. For $A_{\rm v}=1$ km s$^{-1}$, which is close to the
typical granular velocity, the interval between arrivals of
neighboring shocks is near $5$-min (Fig.~\ref{fig:time_sig}, top).
This means that the nonlinear wake in the realistic atmosphere and
for the initial perturbations, corresponding to the solar granular
velocity, exhibits about a $5$-min wave period, i.e. it is longer
than in the linear isothermal case ($\sim 3$-min in the
chromosphere).
%
Time-signature
for
$A_{\rm v}=2$ km s$^{-1}$ shows that the leading shock arrives to the detection point at $t\simeq 190$ s,
while the second shock reaches this
point at $t\simeq 630$ s, i.e. after $\sim 440$ s (Fig.~\ref{fig:time_sig}, bottom panel).
It is clearly seen that the interval between arrival times of two consecutive shocks
is longer for a larger amplitude of the initial pulse.

\section{Discussion and summary}\label{sec:sum}
Frequently observed $\sim 5$-min oscillations in the solar corona
are often explained by leakage of photospheric p-modes along
inclined magnetic field. Vertically propagating acoustic waves with
5-min period are evanescent due to the stratification of the solar
atmosphere as chromospheric acoustic cut-off period is $\sim$ 3 min
(Roberts \cite{roberts}). But, acoustic-gravity waves have a smaller
cut-off frequency when they propagate with the angle about the
vertical. This may allow the photospheric 5-min oscillations to
channel along a stratified chromosphere and penetrate into the
corona (De Pontieu et al. \cite{dep05},  Erd{\'e}lyi et al.
\cite{erd2007}, Fedun et al. \cite{fedun09}). In order to increase
the cut-off period from 3-min up to 5-min, the propagation angle (or
magnetic field inclination) should be $\theta$ $\sim$ 50$^0$
($\cos\theta \approx$ 3/5). Therefore, the leakage of p-modes may
take place only in particular regions of the solar atmosphere, where
the magnetic field is significantly inclined in the chromosphere.

In this paper, we suggest an alternative mechanism to explain the
observed oscillations in the solar corona, which  is based on the
rebound shock model of Hollweg (\cite{hol82}). We numerically solved
the full set of Euler equations for the realistic VAL-C temperature
profile and for a Gaussian velocity pulse launched within the
photosphere. We found that velocity pulses, originating from
granules or magnetic reconnection in the lower regions, lead to
different responses of the chromosphere/transition region than the
periodic acoustic waves resulting from p-modes. It must be
mentioned, that the energy of granular motions is higher than the
energy of p-modes, therefore the impulsively triggered waves should
have more power than those triggered by a periodic driver.

The numerical simulations show that as a result of the rapid
decrease of the equilibrium mass density the initial velocity pulse
quickly steepens into a shock.
The shock propagates into the corona, while the
nonlinear wake is formed in the chromosphere due to the atmospheric
stratification. This nonlinear wake leads to consecutive shocks as
was first shown by Hollweg (\cite{hol82}). The interval between the
arrival times of two consecutive shocks depends on the amplitude of
the initial pulse; a stronger pulse leads to longer intervals. The
initial pulse with the granular velocity of $1$ km s$^{-1}$ leads to
$\sim$ $5$-min intervals between consecutive shocks. Therefore, the
quasi-periodic arrival of consecutive shocks in the solar corona may
cause the intensity oscillations with a period close to the interval
between the shocks.

We implemented a simple 1D analytical model in order to avoid the
propagation of acoustic oscillations with the angle to the vertical.
We showed that purely vertically propagating pulses may lead to
quasi $5$-min oscillations in the corona due to consecutive shocks.
Therefore, it is not necessary to invoke the propagation along
inclined magnetic field in order to explain the observed $5$-min
periodicity in the corona.

One dimensional propagation for acoustic waves is justified for
purely vertical magnetic field. In this case, acoustic waves are in
fact slow magneto-acoustic waves for low plasma $\beta \sim
c_s^2/v_A^2 <1$ and fast magneto-acoustic waves for high plasma
$\beta >1$. Here $v_A$ is the Alfv\'en speed. In the solar
photosphere $\beta$ is larger than unity, but it rapidly decreases
due to the mass density fall off with height (and consequent
increase of the Alfv\'en speed). It becomes smaller than unity in
the chromosphere (Gary \cite{gary2001}) and tends to unity somewhere
between the photosphere and the chromosphere and this surface should
be thinner than the width of the chromosphere. The linear fast and
slow magneto-acoustic waves are coupled near the level of $\beta
\sim 1$ when they propagate obliquely to the magnetic field
(Rosenthal et al. \cite{rosenthal}, Bogdan et al. \cite{bog2003}).
However, these waves remain purely acoustic for parallel propagation
unless the tube dispersive effects are taken into account.
Therefore, a pulse propagating along the vertical magnetic field may
not feel the $\beta \sim 1$ surface. On the other hand, the
acoustic wake, which is formed behind the pulse and oscillates along
the magnetic field, may lead to the non-linear energy transfer into
Alfv\'en waves near the $\beta \sim 1$ region (Zaqarashvili
and Roberts \cite{zaqarashvili}, Kuridze and Zaqarashvili
\cite{kuridze08}). This effect may be revealed as far as one
includes the magnetic field into numerical simulations. Then, a part
of oscillation energy may be transformed into transverse
oscillations near this region, but most of energy will remain in
longitudinal oscillations due to thinness of $\beta \sim 1$ region.
Therefore, $\beta \sim 1$ region may not significantly affect the
formation of rebound shocks in the chromosphere and consequently
quasi-periodic acoustic oscillations in the lower corona. However,
a two-dimensional consideration and inclusion of magnetic
field are necessary for the complete understanding of the
proposed scenario. The first step in this direction was
already done by Murawski and Zaqarashvili (\cite{murawski2010}) in
the modeling of spicule formation, but they considered a
simple temperature profile covering only the
chromosphere-corona. The plasma $\beta$ was considered less
than unity along the whole simulation region, therefore the
effects of the $\beta \sim 1$ region were absent in these
simulations. We intend to consider the rebound shock model in the
case of magnetic field and realistic temperature profile in the
future.

An important consequence of the rebound shock model is that the
interval between consecutive shocks depends on the initial amplitude
of pulse (see Fig. 2). The smaller amplitude pulses lead to shorter
interval between consecutive shocks with lower limit of 3-min, which
is the linear acoustic cut-off period of the solar atmosphere. The
interval is longer for stronger initial pulses being $\sim$ 5-min
for $1$ km s$^{-1}$ in the photosphere. Then the rebound shock model
predicts the longer period oscillations above the regions of strong
granular power. The granulation is suppressed in strong magnetic
field regions of the photosphere (e.g. sunspots), therefore the
initial pulses should have smaller amplitudes there, leading to the
oscillation at near cut-off frequency. Indeed, the strong magnetic
field regions (sunspots, magnetic network cores) show predominantly
3-min oscillations in the chromosphere. Note that some observations
also show the 3-min oscillations in the solar corona above sunspots
(De Moortel et al. \cite{de Moortel02}) and other magnetic
structures (Lin et al. \cite{lin2005}). These oscillations can be
excited as a consequence of consecutive shocks due to chromospheric
3-min oscillations.

As a result, we expect that $\sim 5$-min oscillations can be
detected above the regions where the granular energy is significant
i.e. the quiet Sun and surroundings of active regions/magnetic
network. This is consistent with observations. On the other hand,
the detection of 5-min oscillations in the quiet corona is not an
easy task due to a relatively weak intensity of coronal lines.
However, a careful analysis still can be performed. But, one should
keep in mind that the dynamics of acoustic waves in the field-free
regions and along the vertical magnetic field could be quite
different as it was discussed above. An initial acoustic pulse may
be spread horizontally in field-free regions, while almost the whole
energy would be guided along the field lines in magnetic structures.
Therefore, the amplitude of intensity oscillations could be smaller
in the quiet corona than near active regions and chromospheric
network cores. However, the divergence of magnetic field and
increased thermal conduction may significantly weaken the amplitude
of coronal oscillations, which seem to be quite strong and
non-linear on Figs. 1 and 2. Then, the strong slow wave pulses may
become almost linear once they penetrate into the corona as it is
seen by observations. Our numerical model then requires the
inclusion of magnetic field and the thermal conduction (at least, in
the coronal part of the atmosphere). This will be done in future
studies.

It should be noted that the granular velocities may take values
between 0.5-2 km s$^{-1}$ with peak on 1 km s$^{-1}$. As the
interval between consecutive shocks strongly depends on the
amplitude of the initial pulse, then the resulted coronal
oscillations may take values between 4-7 min with a peak on 5-min.
Indeed, the wavelet analysis of coronal line images obtained by
Hinode/EIS  show that the oscillation power of coronal oscillations
is concentrated at the period in a range of 4-6 min (Wang et al.
\cite{wang09}), which is fully consistent with our theory.

Our simulations were performed for an isolated pulse in order to
show clearly the effect of rebound shocks. However, the solar
photosphere is very dynamic, hence the initial pulse probably is
followed by other pulses. The subsequent pulse coming from the
photosphere may also trigger the consecutive shocks in the
chromosphere. The interaction between rebound shocks formed by
different photospheric pulses may set up complex dynamics in the
chromospheric plasma with immediate influence on the lower corona.
Therefore, the coronal oscillations may have broad spectrum as
we already discussed in the previous paragraph. On the other hand,
if the mean interval between subsequent initial pulses is close to
the mean interval between consecutive shocks, then a resonance may
occur in the chromosphere.  This could be subject of future study.

The excitation of coronal oscillations due to leakage of p-modes may
occur only along significantly inclined magnetic field in the
chromosphere (in order to reduce the cut-off frequency). The
magnetic field lines, which are significantly inclined from the
vertical in the chromosphere, may not reach the corona at all.
Therefore, the p-mode leakage may have problems in real geometry of
active region magnetic field. On the contrary, the rebound shock
mechanism may work in any geometry of magnetic field including the
purely vertical field lines. Therefore, the excitation of 5-min
oscillations in the solar corona by photospheric impulsive drivers
has larger area of application than that of by p-modes. We believe
that the future sophisticated models may shed light on the
excitation of coronal acoustic waves.

Our conclusions are:
\begin{itemize}
\item [(a)] a velocity pulse that is initially launched at the photospheric level (due to granules or reconnection) quickly steepens into a shock and
can penetrate into the corona, while a nonlinear wake that is formed behind this shock leads to consecutive shocks in
the chromosphere;
\item [(b)] for the initial photospheric pulse amplitude of $1$ km s$^{-1}$ the time interval between two
consecutive shocks is $\sim 5$-min; the consecutive shocks propagate upwards and may cause the observed $5$-min intensity oscillations in the solar corona;
\item [(c)] the final conclusion is that the observed $\sim$ 5-min oscillations in the solar corona could be caused by impulsive photospheric
perturbations (convection, reconnection) not necessarily by p-modes.

\end{itemize}
{\it Acknowledgements:}
The authors express their thanks to the unknown referee for his/her
constructive comments. The work of TZ and MK was supported by the
Austrian Fond zur F\"orderung der Wissenschaftlichen Forschung
(project P21197-N16). The work of KM was supported by the Polish
Ministry of Science (the grant for years 2007-2010). TZ was also
supported by the Georgian National Science Foundation grant
GNSF/ST09/4-310. The software used in this work was in part
developed by the DOE-supported ASC / Alliance Center for
Astrophysical Thermonuclear Flashes at the University of Chicago.

\end{document}